\documentclass[a4paper,11pt]{article}
\usepackage{multicol}
\usepackage{bm}
\usepackage{jinstpub}
\usepackage{natbib}

\title{Muon Tomography imaging improvement using optimized scattering tracks data based on Maximum Likelihood Method}%

\author[a]{WANG Xiaodong,}
\author[b,c]{YE Kaixuan,}
\author[d]{WANG Yi,}
\author[a,d]{LI Yulei,}
\author[a]{WEI Xie,}
\author[a]{Luo lingyi,}
\author[a]{and CHEN Guoxiang}

\affiliation[a]{School of Nuclear Science and Technology, University of South China, Hengyang 421001, China}
\affiliation[b]{Institute of Plasma Physics Chinese Academy of Sciences, PO Box 1126, Hefei, Anhui 230031, China}
\affiliation[c]{University of Science and Technology of China, Hefei, Anhui 230026, China }
\affiliation[d]{Tsinghua University, Department of Engineering Physics, Beijing, 100084,China}
\emailAdd{wangxd@usc.edu.cn}%

\abstract{Point of colsest Approche algorithm (PoCA) based on the formalism of muon radiography using Multiple Coulomb scattering (MCS) as information source is previously used to obtain the reconstruction image of high Z material. The low accuracy of reconstruction image is caused by two factors: the flux of natural muon and the assumption of single scattering in PoCA algorithm. In this paper, the maximum likelihood method based on the characteristics of Gaussian-like distribution of muon tracks by MCS is used to predict the optimal track of outgoing muon. The receiver operating characteristic (ROC) and the localization ROC (LROC) are used as two analysis methods to evaluate the quality of reconstruction image. From the results of simulation, the perfect discrimination of longitudinal materials could be well achieved by maximum likelihood algorithm and the discriminate ratio that is predicted by the maximum likelihood method is about 15$\%$ higher than that of predicted by PoCA algorithm method. It is seen that the maximum likelihood method can greatly improve the accuracy of reconstruction image.}


\arxivnumber{1234.56789(???)} % only if you have one

\proceeding{XIV$^{\text{th}}$ Workshop on RESISTIVE PLATE CHAMBERS AND RELATED DETECTORS\\
  02.19-23,2018\\
  Puerto Vallarta, Mexico}

\begin{document}
\maketitle
\flushbottom

\section{Introduction}
Cosmic-ray muon has high penetrability, which has been used for non-destructing imaging for decades by imaging methods ranging from 2-D analysis using measurement of muon attenuation to 3D muon tomography based on Multiple Coulomb scattering (MCS) of muons as they across the materials. The attraction of muon imaging technology is no manufactured radiation source, no artificial dose, and high sensitivity to high Z material ,such as special nuclear material (SNM) refs.~\cite{1}. The scattering information of muon can be recorded by a pair of position sensitive detectors, such as Drift-Tube detector refs.~\cite{1,2}, Gas Electron Multiplier (GEM) refs.~\cite{3,4}, Multi-gap Resistive Plate Chamber (MRPC) refs.~\cite{5,6,7}, which are placed on both sides of the detected object. The Point of Closet Approach (PoCA) firstly proposed by Los Alamos National Laboratory (LANL), is one of the most widely used reconstruction algorithms ,which has been investigated for detection of high-Z material refs.~\cite{8,9,10} in last decades. The low accuracy of reconstruction image is caused by two factors: the low flux of natural muon and the assumption of single scattering in PoCA algorithm.

In this paper, the superiority and ability of the maximum likelihood scattering-expectation maximization (MLS-EM) refs.~\cite{11,12} algorithm,is presented. The characteristics of Gaussian-like distribution of muon tracks by MCS is used to predict the optimal track of outgoing muon. Maximum likelihood is widely used in reconstruction of image, however, a different iterative Expectation Maximization (EM) algorithm is introduced to find the ML estimate of scattering density profiles of material, which is efficient and flexible refs.~\cite{14} in medical image reconstruction. The images of three models of detected material by comparing the qulity of images with the result of closet approach (POCA) is studied. And the quality of image is researched by the ROC (Receiver Operation Characteristic) curve as well as the localization ROC (LROC) using the MATLAB software.

\section{Concept of muon tomography}
\subsection{Tomography for muons}
The principle of tomography is illustrated in Fig.1, where tomography refers to the reconstruction image of object from projections taken from many different directions. $M$ rays sample the object characteristic function when they pass through the imaging area along the line. When the $i^{th}$ ray passes the imaging area, its sampling (or signal) can be observed. The relationship between a ray's sampling and the discrete object characteristic function can be described as following expression:
\begin{equation}\label{eq1}
P_i=\sum_j w_{ij} f_j,
\end{equation}
where, the weight $w_{ij}$ is the path length of the $i^{th}$ ray through the $j^{th}$ pixel or voxel. The reconstructed characteristic function $\bm{\hat{f}}$ can be estimated by solving the system of linear equations \eqref{eq1},.
Since muon tomography is based on the traditional method, there are still several assumptions to be modified:
\begin{itemize}
  \item The cosmic ray muon has a broad angular distribution around zenith with limited flux.
  \item The ray signal, the angle of MCS is, the multiple Coulomb scattering is stochastic with a zero-mean Gaussian, and the actual distribution even has heavier tail.
  \item the rough location could be realized, and the structure of high Z material is estimated roughly..
\end{itemize}

\begin{figure}[htbp]
\centering
\includegraphics[width=8cm]{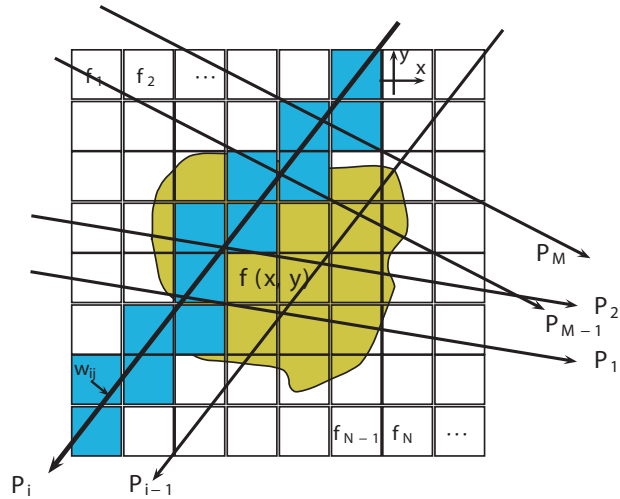}
\caption{(color online) The principle of traditional tomography. A discrete model of the object characteristic function is adopted by assuming uniform values within each pixel or voxel, denoted by the values $f_1,f_2,\cdots,f_N$. And the rays sampling can be observed.}
\end{figure}

As is illustrated in Fig.2, the scattering angle is exaggerated, the observed data $D_i$ of muon is the scattering angle $\Delta\theta_i$. The x and y planes for each of the scattering angle $\Delta\theta_i$ to add reconstructed information is used in muon tomography,  $D_{x,i}~(D_{y,i})$ as:
\begin{equation}\label{eq2}
\begin{split}
D_{x,i}&=\Delta\theta_{x,i}=(\theta_{x,out}-\theta_{x,in})_i\\
D_{y,i}&=\Delta\theta_{y,i}=(\theta_{y,out}-\theta_{y,in})_i.
\end{split}
\end{equation}

\begin{figure}[htbp]
\centering 
\includegraphics[width=8cm]{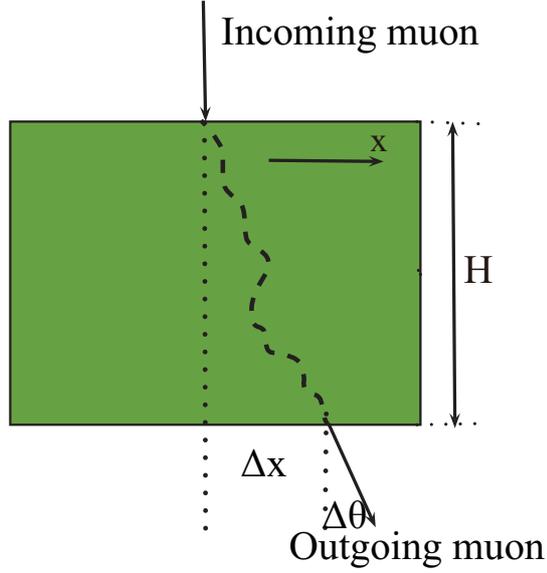}
\caption{(Color online)The trajectory of muon through the material , it can be described by the scattering angle $\Delta\theta$ and displacement $\Delta x$. Here, the path length of ray is approximately equal to the thickness of material $H$.}
\end{figure}

The conditional probability distribution of the observed data $D_i$ may be approximated as Gaussian distribution[9] with a zero mean, which is defined as:
\begin{equation}\label{eq3}
  P(D_{i}|\bm{\lambda})=\frac{1}{\sqrt{2\pi}|\Sigma_i|^{1/2}}exp{\left(-\frac{D_{i}^2}{2\Sigma_i}\right)}.
\end{equation}

Where $\Sigma_i$ is the variance; $\bm{\lambda}$ is the scattering density ditribution.
\par
The variance $\Sigma_i$ can be expressed as
\begin{equation}\label{eq4}
  \Sigma_i =p^2_{r,i}\sum_{j}L_{ij}\lambda_j,
\end{equation}
where $L_{ij}$ is similar to $w_{ij}$, which is the path length of the $i^{th}$ ray through the $j^{th}$ voxel, and $p_{r,i}$ is the momentum ratio which is inversely proportional to $i^{th}$ muon momentum $p_{i}$.
When considering the muon detector noise, the variance is refined as
\begin{equation}\label{eq5}
  \Sigma_i =C_i+p^2_{r,i}\sum_{j}L_{ij}\lambda_j,
\end{equation}
where, $C_i$ is the contribution of the detector noise. The expression of \eqref{eq1}, \eqref{eq4} and \eqref{eq5} have a similar form, while \eqref{eq4} and \eqref{eq5} could express the relationship between the variance of scattering angle and the scattering density for the stochastic signal.

\subsection{Simulation by GEANT4}

In this simulation, the Monte Carlo simulation package GEANT4 is used to simulate the MT spectrometer based on position sensitive detectors. The sensitive area has a volume of $2\times2\times1~m^3$ filled with work gas. To achieve the reconstruction result accordant with practical circumstances, the number of simulated muon is $2.0\times10^5$ corresponding to 5 min of exposure in experiment and the size of material cubes are $10\times10\times10~cm^3$.

Fig.3 shows that a group of position sensitive detectors are located above the object, and one below. The position, angle and momentum of muon will be recorded in these detectors.

\begin{figure}[htbp]
\centering 
\includegraphics[width=8cm]{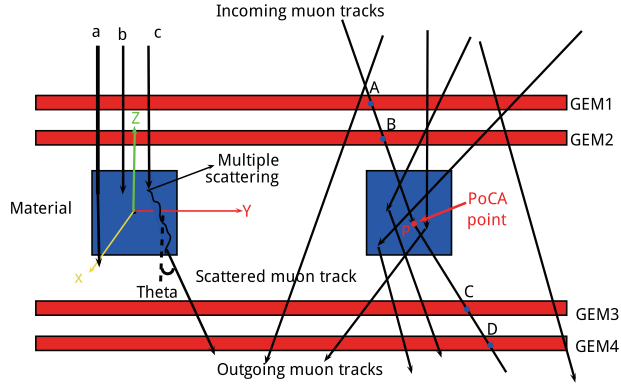}
\caption{(Color online) the principle of muon tomography. The muon has three interactions a, b and c: transmission, energy loss and MCS. The points A,B,C and D are the positions of muon in these detectors respectively, and other information is shown in the figure 3. }
\end{figure}

The physical process of muon, shown in Fig.3, is described in GEANT4 of the physics list QGSP-BERT-HP.The spectrum of generated muon is established by the spectrum of cosmic ray muons in the range of 3 to 100~GeV, which obeys the empirical formula refs.~\cite{14}:
\begin{equation}\label{eq6}
  \frac{dI}{dE\times dcos\theta}=0.14E^{-0.27}\left( \frac{1}{1+\displaystyle\frac{1.1E cos\theta}{115 GeV}}+\frac{0.54}{1+\displaystyle\frac{1.1E cos\theta}{850 GeV}} \right),
\end{equation}
where, $\theta$ is plane angle from vertical and $E$ is the muon energy.

\section{Reconstruction algorithms}
\subsection{PoCA reconstruction algorithm}
Fig.4 shows that the probability density distribution of scattering angle $\Delta\theta$ can be approximated as Gaussian distribution (as the red fitting curve shown) when muons are passing through the aluminum, iron, lead and uranium. The variance of scattering angle is the function of atomic Z number. The results show that the RMS (mrad) are about 4.33 for aluminum, 10.04 for iron, 18.69 for lead, and 25.24 for uranium, respectively.

The PoCA is a simple geometric algorithm under the assumption of single scattering in each event. This algorithm can quickly discriminate the location and structure of object, especially for high Z material, which has been validated through experimental studies refs.~\cite{3,8,9,10,15}.

\begin{figure}[htbp]
\centering 
\includegraphics[width=8cm]{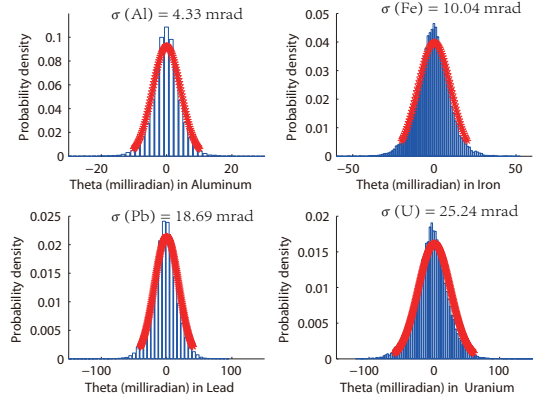}
\caption{(color online) The probability density distribution of scattering angle may be approximated as a zero-mean Gaussian distribution. The RMS scattering are about 4.33 (mrad) for Al (upper-left), 10.04 for Fe (upper-right), 18.69 for Pb (lower-left), and 25.24 for U (lower-right). The blue histogram and red curve are the simulated and fitting results, respectively.}
\end{figure}

Firstly, the scattering angle is in the order of milliradian. The track of each muon can be computed as a straight line connecting the incoming and outgoing points. Secondly, it should be assumed that the scatter of each event only occurred once time on the closest point to the incoming and outgoing tracks, which is called as the PoCA point. But in the view of the three dimensions spatial, the incident and scattered tracks may not be coplanar and not intersect at a point. The point closet to the each pair line (incoming and outgoing) are computed by solving a linear algebraic formulation and the midpoint of common perpendicular are taken as the PoCA point. Fig.5 is the schematic of PoCA algorithm in the voxellation of imaging area.

\begin{figure}[htbp]
\centering 
\includegraphics[width=8cm]{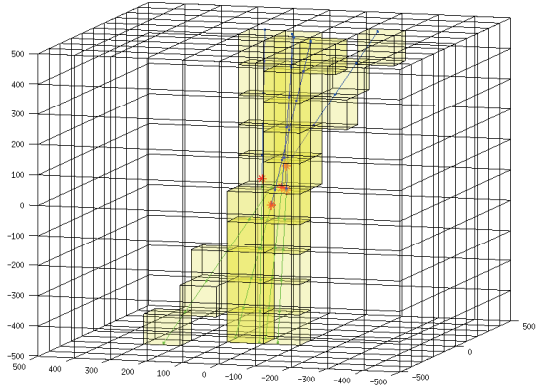}
\caption{(color online) The PoCA algorithm schematic. The blue and green lines are incident and exiting tracks respectively. The yellow cubes are voxels through which the muons pass, and the red points are assumed to be PoCA points.}
\end{figure}

Thirdly, the ray signal can be marked as $s_i$ and defined as:
\begin{equation}\label{eq7}
\begin{split}
  s_i&=\frac{\sqrt{(\Delta\theta^2_{x,i}+\Delta\theta^2_{y,i})}}{2}\\
     &=\left[\frac{(\theta_{x,out}-\theta_{x,in})^2+(\theta_{y,out}-\theta_{y,in})^2}{2}\right]_i.
\end{split}
\end{equation}

Moreover, since muons come from different directions so that path length is different for each muon, we modify \eqref{eq4} to compute the scattering density estimate $\bm{\hat{\lambda}}$ for reconstruction.
\begin{equation}\label{eq8}
  \lambda_j=\frac{1}{p^2_r}\frac{\sigma^2_\theta}{L}=\sum_{i:L_{ij}\neq 0}\frac{s_{ij}^2}{p^2_{r,i}M_jL_{ij}},
\end{equation}
where, the ray signal $s_{ij}$ is:
\begin{equation}\label{eq9}
  s_{ij} = \left\{\begin{array}{rcl}
         s_i& Poca~voxel\\
         0  & along~path~except~Poca~voxel
         \end{array}
         \right
         .
\end{equation}

\subsection{MLS-EM reconstruction algorithm}
The new Maximun Likelihood Scattering-Expectation Maximization (MLS-EM) is designed further by using information of scattering angles, which distribute the scattering location along the ray track instead of assigning to the PoCA point according to  probability statistics. In order to compute the ray path lengths through voxels, the entry points to PoCA points to exit points are connected to estimate.
\par
The total function of muon data may be written as:
\begin{equation}\label{eq10}
P(\bm{D}|\bm{\lambda})=\prod_{i}P(D_i|\bm{\lambda}),
\end{equation}
where the conditional probability distribution $P(D_i|\bm{\lambda})$ is given by \eqref{eq3}.
\par
The MLS estimation of the scattering density $\hat{\bm{\lambda}}$ can be solved by maximizing the log likelihood function:
\begin{equation}
\begin{split}\label{eq11}
  \hat{\bm{\lambda}}_{MLS}&=arg\max_{\bm{\lambda}>\bm{\lambda}_{Air}}LP(\bm{\lambda|D})\\
                     &=arg\max_{\bm{\lambda}>\bm{\lambda}_{Air}}\left(\frac{1}{2}\sum_i\left(-log(\Sigma_i)-\frac{D_i^2}{\Sigma_i}\right)\right),
\end{split}
\end{equation}
an EM algorithm to maximize the log likelihood function are developed
to calculating the scattering density estimate, which is more efficient and flexible than traditional method.
\par
The following MLS-EM update equation for the scattering density at the $n^{th}$ iteration are derived as:
\begin{equation}\label{eq12}
  \hat{\lambda}_{j,MLS-EM}^{n+1}=\frac{1}{2}median_{i:L_{ij\neq0}}B^n_{ij}.
\end{equation}
\par
Since the measurements in x and y are independent, $B_{ij}$ can be computes as the average of $B_{x,ij}$ and $B_{y,ij}$:
\begin{equation}\label{eq13}
  B^n_{x,ij}(B^n_{y,ij})=2\lambda_j^n+\left(\frac{D^2_{x,i}(D^2_{y,i})L_{ij}}{\Sigma_i^2}-\frac{L_{ij}}{\Sigma_i}\right)\times p_{r,i}^2(\lambda_j^n)^2.
\end{equation}
\par
More detailed information can be found in refs.~\cite{9}.
\par

\section{Results and Analysis}
\subsection{Results}
Fig.6 shows the perspective view of three different scenes for comparison between PoCA and MLS-EM algorithms: the horizontal, diagonal and vertical objects. The number of incident muons  is $2.0\times10^5$, corresponding to about 5 mins of exposure in the experiment. The voxels are the size of $50\times50\times50~mm^3$ placed in the imaging area whose size is $2\times2\times1~m^3$.
\par

\begin{figure}[htbp]
\centering 
\includegraphics[width=15cm,height=5cm]{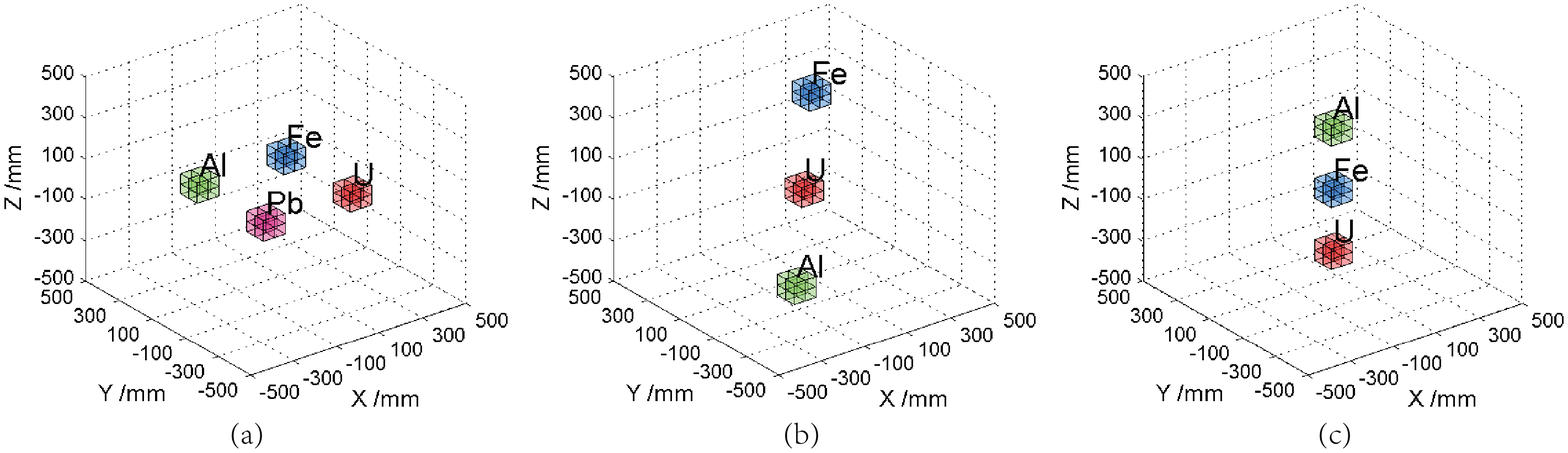}
\caption{(Color online) The perspective view of the simulated horizontal (a), diagonal (b) and vertical (c) scenes. }
\end{figure}

Fig.7 shows the comparison results of horizontal scene between PoCA and MLS-EM. In the reconstructed image, the voxels with scattering density in a range of [-0.5,5) are colored by green (the low-Z), [5,30) blue (the medium-Z), and [30,$+\inf.$] red (the high-Z), respectively. In the MLS-EM reconstruction, the maximum times of iteration is set to 100, and the initial value of voxel scattering density is set to be of air: $\bm{\lambda}^0=10^{-6}$ refs.~\cite{6}. Furthermore, the averages of scattering density in the voxels where objects appear are used to demonstrate the results in brackets. Fig.7(a,c) are the 2D and 3D reconstructed images using PoCA algorithm, respectively, which shows that the scattering density estimate of (Al, Fe, Pb, U) are (1.72, 7.75, 54.63, 57.83 $mrad^2/cm$). Fig.7(b,d) are the 2D and 3D reconstructed images using MLS-EM algorithm, respectively, which shows that the estimation of scattering density  are 1.27, 7.98, 52.68, 62.70 for  Al, Fe, Pb, U,respectively.

\begin{figure}[htbp]
\centering 
\includegraphics[width=8cm]{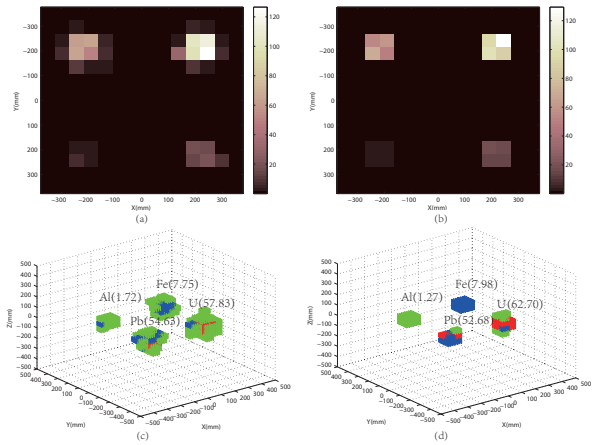}
\caption{(color online) The comparison reconstruction in 2D (a,b) and 3D (c,d) of the horizontal material between PoCA (a,c) algorithm with MLS-EM (b,d) algorithm. }
\end{figure}

Fig.8 displays the comparison results of diagonal scene between PoCA and MLS-EM, respectively, which show that the scattering density of (Al, Fe, U) are (1.36, 7.45, 63.61) by PoCA reconstruction and (1.73, 7.38, 68.37)by MLS-EM correspondingly. Fig.9 shows the comparison results of vertical scene between PoCA and MLS-EM, respectively, which reflect that the scattering density of (Al, Fe, U) are (1.66, 9.96, 58.59) by PoCA reconstruction and (1.53, 8.11, 61.41) by MLS-EM correspondingly.Compared with the PoCA algorithm, the MLS-EM algorithm can reconstruct better location and appearance, particularly in vertical scenes.

\begin{figure}[htbp]
\centering 
\includegraphics[width=8cm]{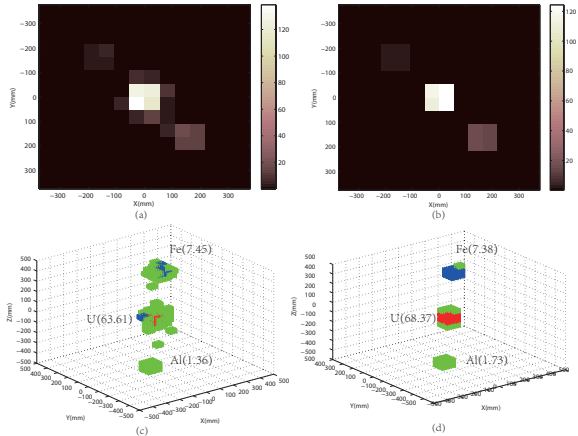}
\caption{(color online) The comparison reconstruction in 2D (a,b) and 3D (c,d) of the diagonal material between PoCA (a,c) algorithm with MLS-EM (b,d) algorithm. }
\end{figure}

\begin{figure}[htbp]
\centering 
\includegraphics[width=8cm]{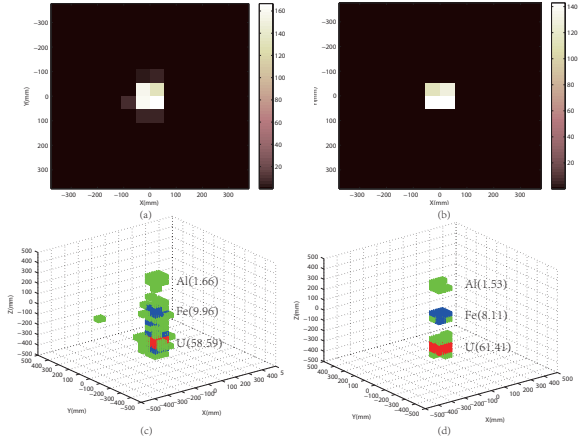}
\caption{(color online) The comparison reconstruction in 2D (a,b) and 3D (c,d) of the vertical material between PoCA (a,c) algorithm with MLS-EM (b,d) algorithm. }
\end{figure}

\subsection{Analysis}
The goal of muon tomography is to discriminate and exclude the presence of high-Z material which can achieve a high efficiency and keeping false positives rate low in a short reconstructed time. We plot the ROC curve and the localization ROC (LROC) curve that has been commonly  used in the binary discrimination system to evaluate the image quality in  PoCA and MLS-EM reconstruction.

\begin{figure}[htbp]
\centering 
\includegraphics[width=8cm]{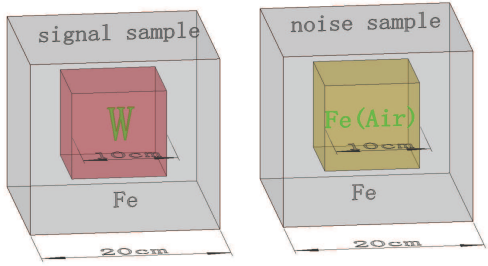}
\caption{(Color online) The ROC (LROC) analysis illustration for muon tomography. }
\end{figure}

Three sets of samples are generated, as are shown in Fig.10. A set of sample hiding the target (a tungsten cube inside the iron volume) is regarded as the "signal" sample, the other two sets of sample without the target (empty or iron cube inside the volume) are regarded as the "noise" samples. The 50 images per set are tested when the event of muon is $8.0\times10^4$ or less. The maximum of scattering density , $\lambda_{max}$, in the image is compared with a selected threshold, $T$. The true positive rate (TPR, sensitivity) is considered as the probability to trigger the alert in the "signal" image when $\lambda_{max}>T$. Otherwise, the false positive rate (FPR, 1-specificity) is defined as the probability to exclude the presence of target in the "noise" image when $\lambda_{max}<T$. A series of pair of sensitive and specificity can be obtained by varying the threshold. The sensitivity plotted against 1-specificity is the ROC curve, the perfect method would yield a point in the upper-left corner of the ROC curve. In that case, the area under the curve (AUC) is equal to 1.

To consider the localization, the LROC curves are also performed, which have been used in the medical imaging community for assessing the lesion localization performance refs.~\cite{12}. The test statistic in LROC differ in that the true positive appears as if and only if the $\lambda_{max}$ is obtained at the target location in the "signal" sample. Fig.11 exhibits the analysis of ROC and LROC curve between PoCA and MLS-EM algorithm. It is clear that PFR, MLS-EM can achieve much higher TPR than PoCA algorithm. For discriminating the signal sample from the noise sample, the MLS-EM reconstruction achieve AUC of 0.9964 in the air noise, 0.9764 in the iron noise for the ROC curve, 0.858 in the air noise, and 0.8592 in the iron noise for the LROC curve, respectively. The results are much higher than those of PoCA reconstruction (0.8708, 0.8366, 0.7312, and 0.798 correspondingly). The comparison reflects that the MLS-EM algorithm can significantly improve the performance of ROC and LROC to the PoCA approach.
\par

\begin{figure}[htbp]
\centering 
\includegraphics[width=10cm,height=8cm]{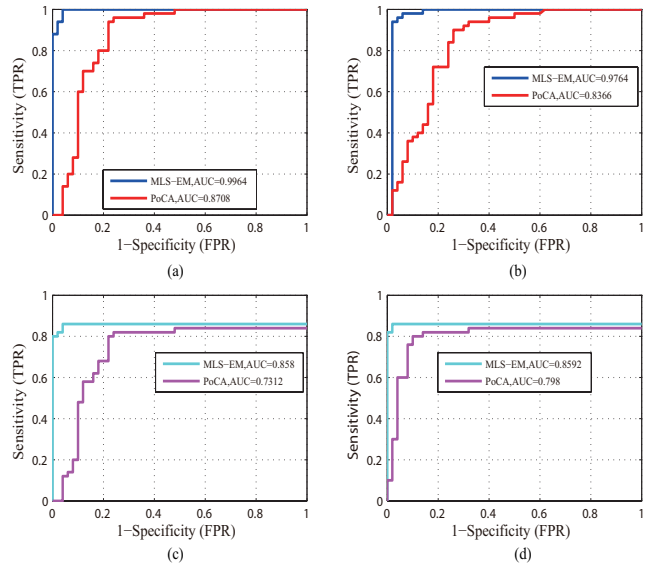}
\caption{(Color online) The analysis of ROC (LROC) and AUC between PoCA and MLS-EM reconstruction: (a) ROC: W vs. Air, (b) ROC: W vs. Fe, (c) LROC: W vs. Air, (d) LROC: W vs. Fe. }
\end{figure}

\section{Conclusion}
This paper describes two tomographic reconstructions based on MCS of natural muons. The PoCA algorithm is employed to detect the high-Z material. And considering the actual application, a instance of EM method is designed to solve large size ML problems which differs and more efficient. The simulated results suggest that, MLS-EM algorithm can significantly improve the imaging performance of muon tomography.

\acknowledgments

This work was supported by National Natural Science Foundation of China (Grant No.11605086), the foundation of Natural Science Foundation of  Hunan Province (Grant No. 2018JJ3422), the  the foundation of Education Department of Hunan Province (Grant No. 15B205), and the foundation of open project of state key laboratory of particle detection and electronics.)


\begin{thebibliography}{99}
\bibitem{1} Borozdin K N, Hogan G E, Morris C, et al, \emph{Surveillance: Radiographic imaging with cosmic-ray muons}, \emph{Nature}. {\bf 422} (2003) 277.
\bibitem{2} Bittner B, Dubbert J, Horvat S, et al,  \emph{Development of fast high-resolution muon drift-tube detectors for high counting rates,}
           \emph{Nucl. Instr. and Meth. A.} {\bf 628} (2011) 154-157.
\bibitem{3} K. Gnanvo, P. Ford, J. Helsby et al,  \emph{Performance expectations for a tomography system using cosmic ray muons and micro pattern gas detectors for the detection of nuclear contraband}, \emph{ Proceeding of IEEE Nuclear Science Symposium Conference Record, Dresden, Germany}. {\bf } (2008) 1278-1284.
\bibitem{4} Jared Sturdy,  \emph{Technical Report CMS-CR-2016-047}, \emph{CERN, Geneva}.{\bf } (2016)
\bibitem{5} Ye J, Cheng J, Yue Q, et al, \emph{Studies on RPC position resolution with different surface resistivity of high voltage provider}, \emph{Nuclear Science Symposium Conference Record, 2008, NSS'08, IEEE}. {\bf }(2008) 917-918.
\bibitem{6}  Baesso P, Cussans D, Thomay C, et al,  \emph{Toward a RPC-based muon tomography system for cargo containers}, \emph{Journal of Instrumentation}. {\bf 9} (2014) C10041.
\bibitem{7} Wang X, Zeng M, Zeng Z, et al,  \emph{The cosmic ray muon tomography facility based on large scale MRPC detectors}, \emph{Nucl. Instr. Meth. A}, {\bf 784} (2015) 390-393.
\bibitem{8} Miyadera H, Borozdin K N, Greene S J, et al,  \emph{Imaging Fukushima Daiichi reactors with muons}, \emph{Aip Advances.} {\bf 3} (2013) 052133.
\bibitem{9}  J. O. Perry,  \emph{Advanced applications of cosmic-ray muon radiography}, \emph{PhD thesis (The University of New Mexico)}.{\bf } (2013)
\bibitem{10} Rhodes C J, \emph{Muon tomography: looking inside dangerous places,} \emph{Science progress.} {\bf 98} (2015) 291-299.
\bibitem{11} AP Dempster, NM Laird, and DB Rubin, \emph{Maximum likelihood from incomplete data via the EM algorithm}, \emph{PJR Stat. Soc. B.} {\bf 39} (1977) 1-38.
\bibitem{12} Green P J, Latuszynski K, Pereyra M, et al, \emph{Bayesian computation: a summary of the current state, and samples backwards and forwards}, \emph{Statistics and Computing.} {\bf 25} (2015) 835-862.
\bibitem{13} Schultz L J, Blanpied G S, Borozdin K N, et al, \emph{Statistical reconstruction for cosmic ray muon tomography,} \emph{IEEE transactions on Image Processing.} {\bf 16} (2007) 1985-1993.
\bibitem{14} Nakamura K, Particle Data Group, \emph{Review of particle physics,} \emph{Journal of Physics G: Nuclear and Particle Physics.} {\bf 37} (2010) 075021.
\bibitem{15} Schultz L J, \emph{Cosmic ray muon radiography}, \emph{PhD thesis (Portland State University)}. {\bf } (2003)

\end{thebibliography}
\end{document}